\documentclass[twocolumn,twoside,slac]{revtex4}
\usepackage{graphicx}
\usepackage{fancyhdr}
\begin{document}

\title{Revison Control in the Grid Era - the unmet challenge}


%
\author{Arthur Kreymer}
\affiliation{Fermilab, Batavia, IL 60510, USA}

\begin{abstract}

As we move to distribute High Energy Physics computing tasks 
throughout the global Grid,
we are encountering ever more severe difficulties installing and selecting
appropriate versions of the supporting products.
Problems show up at every level: the base operating systems and tools,
general purpose utilities like root, and specific releases of application code.
I will discuss some specific examples, 
including what we've learned in commissioning the SAM software for use by CDF.
I will show why revision control can be a truly difficult problem,
and will discuss what we've been doing to measure and control the situation.

\end{abstract}

\maketitle

\thispagestyle{fancy}


\section{Concerns}

I'm here to talk about revision control because this has emerged lately
as an issue of major concern within CDF, especially as we start to
integrate our own code with the SAM data handing system previously
used only at the Fermilab D0 experiment, and as we start using Grid tools
to submit jobs at remote sites.

We have considerable experience distributing the CDF offline 
analysis code on a wide variety of hardware and software platforms,
and in a number of runtime environments 
( online Level 3 triggers, 
online monitoring,
interactive and batch systems. )

When a user comes to us saying "It's Broken !",
we need to know just what it is that they are running.
There can be a surprisingly large number of possibilities
if you stop to count them carefully.
It is important to do so occasionally,
and to take steps to keep this under some level of control.
I will show several specific examples.
 
\section{CDF Releases}

First, let me outline what kinds of offline software releases we support,
just so that the terminology is understood.

Then let's count how many varieties of code we might be
dealing with for a single given line of source code,
in the development release.

This is a worst case, but it's what got me worrying about this problem,
and it is interesting to look at the numbers.

\subsection{Types of Releases}

\begin{itemize}

\item Base

These are fully tested for specified purposes (Production, Analysis )
Version numbers are for example 4.8.0 for general purpose use.

\item Point

Point releases are numbered like 4.8.1, and are otherwise just like
base releases. They have  specific corrections for operations,
and are usually not on the main line of development.
For example, 4.8.4 for Production track restruction for the
2003 winter conferences, and 4.9.1hpt2, for physics analysis for the
same conferences.

\item Integration

Integration releases are (bi)weekly frozen releases which have much less
testing than base release.
We only require that the code compile and link for integration releases.
They provide a stable base against which major code development is conducted.

\item Development

We build a release using the latest source code from the head of the CVS
code repository each night, using the code present at midnight.
We also perform rudementary code execution validition nightly.
This gives early feedback in case of broken code or global problems.

\item Patch

Inevitably there are slight administrative code changes need in the Level3,
online monitoring, and production farm operations which do not merit a formal 
code release, but which need to be reproducible from CVS tags.
Patch releases are a formalization of the usual developers temporary 
working area, using a standard creation script and named patch list.

\end{itemize}

You can find a current list of base, point and integration releases on the web
\footnote{http://cdfkits.fnal.gov/DIST/logs/rel/Releases.html} .

\subsection{Development 2001}

Here we count the different ways in which a given line of code
might have been compiled in the nightly development release, in 2001 .

\begin{itemize}

\item 5 - OS

We were building with Linux 2.0, Linux 2.2, IRIX , OSF/1 and SunOS

\item 3 - Optimization

We built variously with default, minopt (K1) and maxtop (K3)-

\item 3 - History

Some systems do full clean rebuilds daily.
Some systems force a full clean rebuild every Sunday.
Most do daily incremental builds, and rarely if ever do full rebuilds.
These three different histories could leave stale code in some libraries,
as gmake does not know how to clean up when source code is removed.
We have ways of forcing global rebuilds when we think it is needed.
This is not usually a problem, but we have to worry about it nevertheless.

\item 3 - Languages

You might find code built with with the KAI KCC 3, KCC 4 or gcc compilers.

\item 4 - database 

There was automatically generated code for connecting to
text, mSQL, Oracle OCT or Oracle OTL.
So any calibration database code would exist in for different forms.

\item 2 - Library mode

The libraries are mostly built for static linking.
But when for rapid development, 
developers often exercise an option to make local shared object libraries.

\item 2 - build style

Through early 2001, we had only built libraries one at a time,
using a single processor.
Then we transitioned, on the large central systems,
to doing parallel rebuilds.
This is not trivial, as part of the gmake process involved generating
headers and code which need to be subsequently used by other packages.
We think we have this correctly set up now, but this was less obvious
in 2001.

\end{itemize}

The net result of these possibilities is that a given single line of code
in the development release could be built 2160 different ways.

It is clear impossible to exercise and test all of the possibilities
with our finite resources. 

We survived because very few of the options were exercised by normal users.
Most developers ran under Linux 2.2 or IRIX, with default optimization,
with weekly clean rebuilds, KAI 3, using the Oracle OCI interface,
with static libraries and using the standard distributed libraries
built in parallel. So the IRIX/Linux choice was the most commmon issue.

But remember that the team of people supporting the code do have to be
aware of all 2160 variations. 
There were people exercising all the options listed above, 
in various combinations.  
It is essential to track down just what a developer is doing before
starting to investigate a problem.

\subsection{Development 2003}

The situation was not just a relic of rapid transitions during 2001.

Fewer people work out of the development release,
now that Integration releases are available.
So these choices are exercised less often by regular developers.

\begin{itemize}

\item 3 - Operating System

Linux 2.2, Linux 2.4, IRIX

We have dropped OSF1 and SunOS, and will probably drop IRIX soon.

\item 3 - Optimization

Debug (K0), default (K1) and maxopt

\item 5 - Database

text, mSQL, Oracle OTL MySQL, ODBC

It looks like we have added more options here.
We'd like to cut back to text plus one ODBC type API to all
the back end databases, reducing this to 2 varieties of code.

\item 3 - History

clean, weekly clean, stale just as before.

\item 2 - Languages

KAI KCC 4, gcc

We expect to drop KCC, moving to gcc 3.2+ before 2004.

\item 2 - Codegen

We are testing a new simplified code generation structure at the moment.
Eventually this will allow the single database API, making things better.
A this moment, though, it creates one more option to track.

\end{itemize}

There are still a net 810 varieties possible for each line of code.
Again, most are not exercised.
Almost all development is now on Linux 2.4, default optimization,
Oracle OTL, using the Integration base release, with KCC 4,
and the old codegen.
The parallel versus serial builds are now considered equivalent.

\subsection{Product versions}

So far we have just discussed how many ways a single line of code
could get built within the CDF offline release.
That release itself depends on many external products,
which we do not build as part of the release.

We try to take strong control of these versions,
and support only a single specific version of each
for a given CDF offline code release.

A current snapshot is given in Table ~\ref{products}

\begin{table}[t]
\begin{center}
\caption{Product versions}
\begin{tabular}{|l|c|}
\hline \textbf{Product} & \textbf{version}
\\
\hline cdfdab            & v1\_0                   \\       	
\hline cern              & 2000                    \\ 
\hline csl\_offline      & v3\_1                   \\ 
\hline dcap              & v2\_24\_f0203           \\
\hline dddebugger        & v3\_3\_1                \\
\hline gdb               & v5\_0b\_external        \\
\hline gsl               & v1\_2                   \\       	  
\hline gtools            & v2\_4                   \\       	  
\hline herwig            & v6\_4a                  \\        
\hline isajet            & v7\_51a                 \\        
\hline java              & v1\_2\_2b               \\         
\hline kai               & v4\_0f\_a               \\         
\hline lund              & v6\_203a                \\  
\hline level3\_offline   & v2\_1                   \\        
\hline msql              & v2\_0\_10               \\ 
\hline oracle\_client    & v8\_1\_7\_lite          \\  
\hline pdf               & v8\_04                  \\
\hline perl              & v5\_005                 \\
\hline perl\_dbd\_oracle & v5\_005\_817            \\
\hline root              & v3\_05\_00a -q KC\_4\_0 \\        
\hline stdhep            & v4\_09		   \\	       
\hline tcl               & v8\_3\_1a	           \\	 
\hline cdfdb\_data       & v0\_external            \\	       
\hline geant             & v3\_21\_13\_external    \\	       
\hline qq                & v9\_1b\_external	   \\
\hline
\end{tabular}
\label{products}
\end{center}
\end{table}

There are also a small number of products for which
a fixed version is not appropriate, 
where we need to use the 'current version.

These products provide overall infrastructure support (UPS, UPD),
or provide services not tied to any single release (diskcache\_i, kai\_key).

The development and current product versions are updated by 
requiring each system to run a nightly update of a 
'development lite' release of the CDF code,
containing a few standard maintenance and operations scripts,
database access configurations, other small operations data files.

\section{SAM versions}

Having this experince in mind, it was natural when we started to
integrate the SAM data handling software with the CDF offline code
to look at similar issues.

Just as we did when looking at varieties of code in the CDF development
releases, we can look at varieties of operating environments for the
SAM code. We encountered about 60 configuration files,
used by a variety of programs, tailored for each of the following :

\begin{itemize}

\item 2 - Experiments

D0 or CDF

\item 3 - Databases

development, integration, or production

\item 5 - Operating systems

IRIX, Linux, OSF/1, SunOS and generic

\item - usage

User, Station, Stager, CORBA server

\end{itemize}

This makes 120 varieties to consider, clearly excessive.

After some investigation, 
we found that the D0 experiment does not use these configuration files.
CDF runs SAM clients only under Linux,
and does not want the option of different product versions
for any of the options listed above.
So in this case, we can reduce the complexity
by simply removing the options entirely.

I have mentioned this small example mainly because this is what immediately
motivated the presentation of this talk.

\section{Java}

There is the temptation to look to portable virtual environments
such as Java for a solution to portability and revision control problems.
So I'll deliver a little word of warning that things are not so simple.

At this time there are at least five versions of java used in the CDF offline
environment, each somewhat incompatible with the other in function or licensing.
I'm sure that even more versions are in use in the online system.

\begin{itemize}

\item v1.1.6 on IRIX
\item v1.2.2b on Linux
\item v1.1.7 was used on SunOS and OSF1 before we retired them
\item v1.3 used for dcache data handling software development
\item v1.4 used for dcache deployment in production at Fermilab

Even in the very simple offline application,
where we use Java to convert data struction descriptions into
C++ calibration database interface code (reading and writing plain text),
we have had severe portability problems, particularly on SMP systems.
The java application will die with strange and inappropriate diagnostics,
something we've never really been able to find a cause for.
At the moment we seem to be able to work around this by setting
environment variable LD\_ASSUME\_KERNEL=2.2.5 ,
 
\end{itemize}

\section{Operating systems}

It is tempting when faced with operating system release issues
to demand that everyone run the same system.
But CDF is a worldwide collaboration. 
Most of the systems are not under Fermilab control,
and are subject to legitemate local constraints.

Here is a little snapshot of RedHat releases, and kernels
in use within CDF on 22 March 2003.

The hundreds of centrally managed systems in the
Production farms and the Central Analysis Farm batch system
all run Fermi Linux 7.3 .

The office desktop systems in Table ~\ref{desktop} 
and offsite systems in Table ~\ref{offsite} are a very mixed bag.
Note that the offsite survery only lists systems I can log into,
and does not include any of the large offsite farm systems.

\begin{table}[t]
\begin{center}
\caption{CDF Desktop Fermi Linux deployment}
\begin{tabular}{|r|c|}
\hline \textbf{systems} & \textbf{Fermi Linux}
\\
\hline 64   & 6.1.1	   \\
\hline 33   & 6.1.2	   \\
\hline 121  & 7.1.1	   \\
\hline 67   & 7.3.1a	   \\
\hline
\end{tabular}
\label{desktop}
\end{center}
\end{table}

\begin{table}[t]
\begin{center}
\caption{CDF off site Linux systems}
\begin{tabular}{|r|c|l|}
\hline \textbf{systems} & \textbf{type} & \textbf{Linux}
\\
\hline  1  & RedHat & 6.1    \\
\hline  6  & Fermi  & 6.1.1  \\
\hline  2  & RedHat & 6.2    \\
\hline  7  & RedHat & 7.1    \\
\hline 11  & Fermi  & 7.1.1  \\
\hline  5  & RedHat & 7.2    \\
\hline  3  & Fermi  & 7.3.1  \\
\hline
\end{tabular}
\label{offsite}
\end{center}
\end{table}

There are 21 different Linux kernels in use offsite,
each on 1 to 4 nodes, as shown in Table ~\ref{kernel}

Therefore, we try very hard, 
and change the code if necessary,
to keep the CDF offline code independent of the kernel.

\begin{table}[t]
\begin{center}
\caption{CDF offsite kernels}
\begin{tabular}{|r|l|}
\hline \textbf{systems} & \textbf{kernel}
\\
\hline 1 & 2.2.16-3		    \\
\hline 1 & 2.2.16-3-ide 	    \\
\hline 4 & 2.2.16-3smp  	    \\
\hline
\hline 1 & 2.2.17-14		    \\
\hline
\hline 1 & 2.2.19-6.2.1 	    \\
\hline 1 & 2.2.19-6.2.16enterprise  \\
\hline
\hline 3 & 2.4.3-12		    \\
\hline 1 & 2.4.3-12smp  	    \\
\hline
\hline 2 & 2.4.5		    \\
\hline 1 & 2.4.5-4G-rtc 	    \\
\hline
\hline 1 & 2.4.7-10		    \\
\hline 2 & 2.4.7-10smp  	    \\
\hline
\hline 1 & 2.4.9-12smp  	    \\
\hline 1 & 2.4.9-21smp  	    \\
\hline 2 & 2.4.9-34		    \\
\hline
\hline 3 & 2.4.18		    \\
\hline 2 & 2.4.18-10smp 	    \\
\hline 3 & 2.4.18-19.7.xsmp	    \\
\hline 1 & 2.4.18-5bigmem	    \\
\hline 1 & 2.4.18RUI2S3 	    \\
\hline
\hline 2 & 2.4.19		    \\
\hline
\end{tabular}
\label{kernel}
\end{center}
\end{table}

\section{root}

Even for a single release of a single product, such as root,
there can be many varieties when faced with the requirements of
various experiments just at Fermilab.
We presently build 15 varieties of each release of root,
listed in Table ~\ref{root} .
We are working hard to reduce this work load,
by moving to a common gcc compiler, and moving to enable C++
exception handling in the CDF code.

Note that the UPS/UPD tools for selecting and using just the
desired version of a product is essential for survival
If a given system can only have one version of root in use,
shared by all projects, then most users will be unable to use it.

\begin{table}[t]
\begin{center}
\caption{Fermilab varieties for root 3.05.03b}
\begin{tabular}{|c|l|}
\hline \textbf{Flavor} & \textbf{Variety}            \\

\hline Linux+2.2  &  GCC\_3\_1 opt		     \\
\hline Linux+2.2  &  GCC\_3\_1  		     \\
\hline Linux+2.2  &  KCC\_4\_0 opt		     \\
\hline Linux+2.2  &  KCC\_4\_0  		     \\
\hline
\hline Linux+2.4  &  GCC\_3\_1 exception opt thread  \\
\hline Linux+2.4  &  GCC\_3\_1 opt		     \\
\hline Linux+2.4  &  GCC\_3\_1  		     \\
\hline Linux+2.4  &  KCC\_4\_0 exception opt thread  \\
\hline Linux+2.4  &  KCC\_4\_0 opt		     \\
\hline Linux+2.4  &  KCC\_4\_0  		     \\
\hline
\hline IRIX+6.5   &  GCC\_3\_1 exception opt thread  \\
\hline IRIX+6.5   &  GCC\_3\_1 opt		     \\
\hline IRIX+6.5   &  GCC\_3\_1  		     \\
\hline IRIX+6.5   &  KCC\_4\_0 exception opt thread  \\
\hline IRIX+6.5   &  KCC\_4\_0 opt		     \\
\hline
\end{tabular}
\label{root}
\end{center}
\end{table}
 
\section{File systems}

Likewise, there are many filesystems in use,
and one hopes that none of the offline physics code
depends on these systems.
This is not a trivial task.

Under Linux, one can use ext2, ext3, xfs and reiserfs
transparently.

There is also limited use of afs, but without Fermilab support.
So we try to avoid doing things that would break thisafs usage.

On the other hand, file systems like NTFS which do not distinguish
upper and lower case characters in file names cannot be used safely,
and we will not try to 'fix' this in the CDF code.

\section{Grid issues}

It is hard to quantify the complexity of the Grid space,
as the tools to be used are just emerging.
Some of the choices  will involve :

\begin{itemize}

\item Authentication

Who is the user requesting services ?
There are a variety of methods in use outside the Grid
( kerberos 4 and 5, ssh, etc.)
Various Grid certificate authorities have been commissioned,
and are in negotiations right now over issues of mutual trust
and subsequent use of these certificates for authorization.

\item Authorization
What is the user allowed to do.
The classic issues are file access (as in afs), job submission/login,
and file copying. In the Grid this is generalized to a broader range
of services, hopefully with a single integrated set of authorization tools.

\item Batch systems
The primary initial use of Grid tools seems to be to access batch
computing facilities. There are a large number of local batch system,
(lsf, pbs, fbsng, etc.),and effort is underway to make this transparent
to the end user.

\item Database access
It take more to run a job these days than just handline input, output,
and the executeable environment. Database access to a central server
is ofter requires, raising issues of network performance, or cacheing
when multiple tier servers are involved.

\item Multi experiments
It is quite nontrivial to manage the environment for even one experiment.
In the Grid world we are expected to share the same environment.
This will require careful attention to detail to see that
undesired couplings do not arise.

\end{itemize}

\section{Conclusions}

A few simple items of advice seem to have emerged

\begin{itemize}

\item Count the varieties or configurations, and pay attention.

\item Reduce cross-project couplings

\item Reduce sensitivity to revisions (file systems, kernels, etc)

\end{itemize}

\end{document}